# Cross-Subject Depression Level Classification Using EEG Signals with a Sample Confidence Method

Zhong-Yi Zhang ⁺⁺, Chen-Yang Xu ⁺⁺, Li-Xuan Zhao, Hui-Rang Hou, Qing-Hao Meng

⁺⁺ Co-first authors with equal contributions

*Abstract*—Electroencephalogram (EEG) is a non-invasive tool for real-time neural monitoring, widely used in depression detection via deep learning. However, existing models primarily focus on binary classification (depression/normal), lacking granularity for severity assessment. To address this, we proposed the DepL-GCN, i.e., Depression Level classification based on GCN model. This model tackles two key challenges: (1) subjectivity in depression-level labeling due to patient self-report biases, and (2) class imbalance across severity categories.

Inspired by the model learning patterns, we introduced two novel modules: the sample confidence module and the minority sample penalty module. The former leverages the L2-norm of prediction errors to progressively filter EEG samples with weak label alignment during training, thereby reducing the impact of subjectivity; the latter automatically upweights misclassified minority-class samples to address imbalance issues.

After testing on two public EEG datasets, DepL-GCN achieved accuracies of 81.13% and 81.36% for multi-class severity recognition, outperforming baseline models. Ablation studies confirmed both modules' contributions. We further discussed the strengths and limitations of regression-based models for depression-level recognition.

*Index Terms*—Depression level recognition, EEG, minority sample penalty, sample confidence.

## I. Introduction

MAJOR depressive disorder (MDD) is a widespread mental illness characterized by sustained sadness, diminished pleasure, lack of enjoyment and, in severe cases, suicidal thoughts [1], [2]. The World Health Organization (WHO) reported in 2019 that over 280 million individuals are affected by MDD worldwide [3]. This condition represents a significant challenge to worldwide public health systems and places a substantial economic strain on global communities and various societal sectors [4]. Consequently, there is an immediate need for efficient methods of diagnosing and treating MDD.

Traditionally, the diagnosis of depression relies heavily on clinical interviews with patients and the administration of psychological questionnaires [5]. However, these methods can be limited by the clinician's skills in the former case, and by the patient's subjectivity, reluctance to disclose, or tendency to conceal symptoms in the latter [6]. As a result, there is a significant research effort dedicated to creating objective measures for assessing depression to avoid the complexity of clinical diagnosis [7].

EEG has gained prominence as an economical and non-invasive tool with superior temporal precision, positioning it as a prime candidate for studying brain activity [8]. This has prompted a number of innovative studies to explore the use of EEG signals to support the diagnostic process of depression. Initial investigations into EEG for depression have involved the combination of manual feature extraction with machine learning techniques [9], [10], [11]. However, due to the complexity of EEG signals and individual differences among subjects, there is still much room for improving the accuracy of these diagnostic methods. The advent of deep learning has spurred its application in depression classification [12], [13], [14], [15], where its powerful feature extraction and pattern recognition capabilities improve diagnostic accuracy.

At present, one can divide the current deep learning models for depression recognition into two categories: binary-classification models, which output only "yes" or "no" (i.e., "depressed" or "healthy"), and multi-classification models, which output different levels (usually categorized into four or five). To the best of the authors' knowledge, most of the models in existing studies fall into the binary classification, and there has been very limited research on models that utilize EEG signals to identify levels of depression.

The most intuitive difference between the task of identifying depressive levels and the binary depression is that the dimensions of the model's final output are different. However, a careful study reveals that migrating the model from the binary to multiple outputs will face the following challenges: First, the results of categorizing the level of depression can be affected by the subjectivity of subjects. In the binary classification task, the subject's depression state is clearly defined, and both patients with depression disorders and healthy subjects are identified after professional questionnaires and assessments by specialized physicians. However, in the depression-level recognition task, the depression levels of patients with depressive disorders in the dataset were not assessed by professional doctors, so the depression level labeling process can only rely on the depression assessment questionnaire scores of the subjects provided in the dataset. Although the scores of these professional questionnaires can reflect the depression levels of

Manuscript received xxxx; revised xxxx; accepted xxxx. Date of publication xxxx; date of current version xxxx. This work is supported by the National Natural Science Foundation of China under Grant 62203321, in part by the China Postdoctoral Science Foundation under Grant 2021M692390. Recommended for acceptance by xxx. *(Corresponding author: Qing-Hao Meng).*

Zhong-Yi Zhang, Chen-Yang Xu, Li-Xuan Zhao, Hui-Rang Hou and Qing-Hao Meng are with the Tianjin Key Laboratory of Process Measurement and Control, Institute of Robotics and Autonomous Systems, School of Electrical and Information Engineering, Tianjin University, Tianjin 300072, China (email: zy_zhang_auto@tju.edu.cn; xuchenyang@tju.edu.cn; zhaolx.tju@gamil.com; houhuirang@tju.edu.cn; qh_meng@tju.edu.cn ).

patients with depressive disorders to a large extent, the depression assessment questionnaires are inevitably subject to the subjective influence of the patients, resulting in a certain gap between the final assessment results and the actual depression levels. Secondly, the sample size of different depression-level categories is unbalanced. In the binary classification task based on the MODMA (multi-modal open dataset for mental-disorder analysis) [16] and PRED+CT (patient repository of EEG data + computational tools) [17] datasets, there is only a slight difference in the number of patients with depressive disorders and healthy subjects, which does not have much impact on the training process of the model. However, after migrating the model to the multi-classification task of depression levels, there is a significant difference in the number of subjects with different depression levels. The imbalance of samples will seriously affect the training of the model. If it is not restricted, the model will tend to predict all samples as the category with the largest number of samples.

To address the above problems, we proposed appropriate solutions for each of them. First of all, with regard to the issue of the possible subjective influence of subjects on the results of depression-level classification, from the data perspective, this issue can be understood by the fact that the EEG data of some of the subjects in the dataset are not fully consistent with the depression-level labels. Therefore, finding these samples with loose correspondences and reducing the learning of these samples by the model is the core idea to solve this problem. To address this problem, we proposed a sample confidence module. During the training process, this module can dynamically evaluate the confidence of the current subject sample according to the stage results of the training, and adjust the model learning degree of the sample in subsequent training according to the size of the confidence, so as to reduce the impact of the samples with poor correspondence between data and label on model training, and thus solving the problem that depression-level labels may be affected by subjective judgment. As for the problem of sample size imbalance, we added a minority sample penalty module to the model. During the training process, this module can increase the model's attention to the subjects in the minority sample category, thereby affecting the parameter update path of the model, so that the parameter update direction fits as many samples in these minority sample categories as possible, thus improving the model's performance in the minority sample category. Finally, on the basis of the basic model, after introducing relevant modules designed for the depression-level classification task, we proposed the DepL-GCN model.

The main innovations and contributions of this paper are as follows:

- We attempted an EEG-based depression-level classification task and succeeded for the first time in migrating a binary-classification model of depressive disorders to a multi-classification depression-level task.
- We proposed a sample confidence module. This module can effectively reduce the model's attention to samples with poor correspondence between data and labels, effectively mitigating the disadvantage of depression-level labels being subjectively influenced by subjects.
- A minority sample penalty module is proposed. In the absence of a dataset specifically for depression level recognition, this module can effectively solve the negative impact of sample imbalance on the model.
- The proposed DepL-GCN model has been extensively verified by experiments on two datasets, and the experimental results validate the effectiveness and superiority of the model.
- For the first time, a depression-level recognition scheme based on EEG and regression model is discussed.

## II. RELATED WORK

Compared with the binary classification of depressive disorders, relatively few studies related to the identification of depression levels have been conducted so far. If the recognition signals are further limited to EEG, there are even fewer related studies. In view of this, in order to introduce the current research status of depression-level recognition models, we have statistically analyzed the research using other physiological signals and deep learning methods, such as using speech signals and facial representations of subjects to identify the state of patients with depressive disorders.

In a study on using EEG signals to identify depression levels, Mohammadi *et al.* [18] used the functional connectivity and complexity of EEG signals to predict the Beck Depression Inventory (BDI) [19] score, which represents the severity of depression. The team conducted cross-subject experiments on private datasets using feature extraction, functional connectivity analysis, and linear regression methods with leave-one-out cross-validation, and finally achieved experimental results with a mean absolute error (MAE) of 6.11 and a root mean square error (RMSE) of 7.69. Another study by Mohammadi *et al.* [20] used different machine learning methods including support vector machine (SVM) and feedforward neural networks to identify depressive disorders and depression levels on EEG signals, and combined different features such as FuzzyEn, FuzzyFractal, and KATZ. This study directly migrated the binary-classification model of depressive disorders to the multivariate classification of depression levels without any special treatment. Under the condition of combining all features, the model achieved an accuracy of about 65% for quadruple classification of depression levels on private datasets (specific data not given in the original text). Mahato *et al.* [21] used six-channel EEG data to detect depression and assess the severity of depression. The team used the SVM classifier and the ReliefF method to select six channels (FT7, FT8, T7, T8, TP7, TP8) from the EEG data for feature extraction. In the four-category task, the SVM classifier achieved an accuracy of 79.17%. Zeng *et al.* [22] used EEG power spectrum and functional connectivity methods to analyze 65 patients with late-life depression and 40 healthy elderly participants to investigate cognitive impairment in elderly depressed patients. The study found that patients with moderate to major late-life depression (LLD) have impaired attention and executive functioning and enhanced power in the posterior upper gamma band. The focus of this study was to understand the neurobiological basis of LLD, and the model was not used to directly identify levels of depression. In a study published in 2024, Zhang *et al.* [23] first used EEG signals and graph convolutional neural (GCN) network for depression level recognition. In order to solve the problems of large inter-individual differences and imbalance of sample size among different categories in the multi-classification problem of depression levels, domain generalization method and minority sample penalty modules were added to the GCN network, respectively. Experimental verification was carried out on MODMA and PRED+CT datasets, with classification accuracies of 75.47% and 77.97%, respectively.

In the study of depression-level recognition based on other physiological signals, Niu *et al.* [24] proposed a time-frequency channel attention and vectorization (TFCAV) network in 2022, which automatically predicts depression levels with facial representation of subjects as input signals. The TFCAV network structure includes convolutional layers, dense blocks, and transition layers. On the AVEC2013 dataset, the study achieved an RMSE of 8.15 and an MAE of 6.01. On the AVEC2014 dataset, the minimum RMSE and MAE of the model were 8.96 and 7.00, respectively. The experimental results are better than the current research work on depression-level classifications using EEG signals. Another study by Niu *et al.* [25] introduced a system called "Depressioner" that uses facial dynamic representation to automatically predict depression levels. This study compared the effects of handcrafted features and convolutional neural network (CNN)-based automatic feature extraction methods on model results. Finally, on the AVEC2013 dataset, the CNN-based feature extraction method showed lower RMSE and MAE, which were 7.41 and 6.09, respectively. In the field of speech prediction of depression levels, Li *et al.* [26] proposed a hybrid network model, which evaluates the depression-level prediction performance of different network architectures on the dataset through long-term global information embedding technology. The results showed that the hybrid network model combining long short-term memory (LSTM) and CNN with channel attention mechanism achieved the best prediction effect. On the AVEC2013 and AVEC2014 datasets, the RMSE and MAE of the model were 7.65/6.01 and 7.32/5.39, respectively, showing lower errors compared with other traditional models.

The above studies show that it is feasible to identify depression levels through EEG signals and other physiological signals, and with the development of deep learning technology, the accuracy of prediction is gradually improving.

## III. METHODOLOGY

In this section, we first introduce the basic model we used, SSPA-GCN, and then describe the overall framework of the proposed DepL-GCN model. Next, we explain in detail the sample confidence module and the minority sample penalty module in the DepL-GCN model. Finally, we experimentally validate the effectiveness of the proposed model and explore other possible implementations of the depression-level recognition task.

### A. Baseline Model

We use the SSPA-GCN model proposed by Zhang et al. [27] as the base model. The SSPA-GCN model, a deep learning architecture based on graph convolutional networks (GCNs), is designed for the task of identifying depression disorders based on EEG signals.

The model combines a multi-dimensional attention mechanism and an improved domain generalization method. By introducing a two-dimensional attention matrix at the input of the model, it can simultaneously focus on the importance of different EEG channels and their internal frequency band features. In addition, SSPA-GCN optimizes the generation of domain labels through a secondary subject partitioning (SSP) strategy, which groups subjects with similar data distribution into the same subdomain, thereby reducing the number of subdomains and increasing the amount of data per subdomain, thus improving the performance of the domain generalization module. This improvement enables the model to perform well in cross-subject recognition tasks, effectively capturing common features across individuals and improving the accuracy and generalization ability of depressive disorder recognition.

In terms of data preprocessing methods, we adopt the same processing method as the SSPA-GCN model. We slice the data at 2-second intervals and extract the differential entropy [28] features of the five frequency bands as inputs to the model, rather than using the original EEG signals directly.

The framework of the SSPA-GCN model and the data preprocessing method are shown in Fig. 1 and Fig. 2, respectively.

### B. Overall Framework

The overall framework of the DepL-GCN model is shown in Fig. 3.
Based on the SSPA-GCN model, we proposed a sample confidence module and a minority sample penalty module to address the difficulty of the depression-level recognition task. In terms of data preprocessing, the DepL-GCN model adopts the same data processing method as the SSPA-GCN model. After the input data passes through the model and the predicted depression level is obtained, the subject's *NeL$_2$* (New *eL$_2$* norm) is calculated based on the predicted value, and then the confidence coefficient is assigned to the sample based on the *NeL$_2$* value. At the same time, if the sample belongs to the minority sample category and the prediction is wrong, we assign the minority sample penalty coefficient to the subject based on *NeL$2$*. Before the model back-propagates and updates the parameters, we redistribute the weights of the samples in the loss function according to the calculated confidence coefficient and minority sample penalty coefficient, which affects the model to increase or decrease its attention to these samples.

### C. Sample Confidence Evaluation Module

According to the above introduction, the sample confidence module proposed in this study is mainly used to solve the problem that the results of depression-level classification may be affected by the subject's subjective influence. From the perspective of model training and data, this problem can be abstracted as follows: there are subjects in the dataset whose EEG data do not strictly correspond to the depression levels based on questionnaire scores. The core idea of solving this problem is to find these poorly corresponding samples and make them play as small a role as possible in the model training process, i.e., to control the gradient information of the samples so that they have less impact on the model parameter update. To this end, a new model learning strategy is proposed to solve the above problem by combining the learning characteristics of the model.

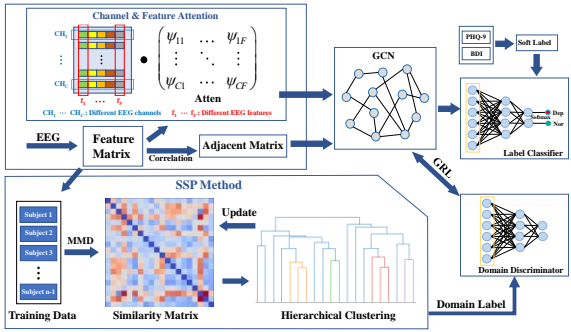

**Fig. 1.** The overview framework of the SSPA-GCN model [27].

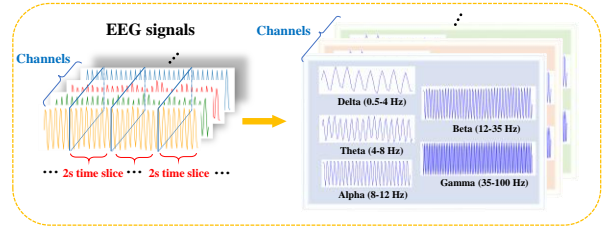

**Fig. 2.** Data preprocessing method. The EEG data is sliced into 2-second slices, and differential entropy features for five frequency bands are extracted as inputs to the model.

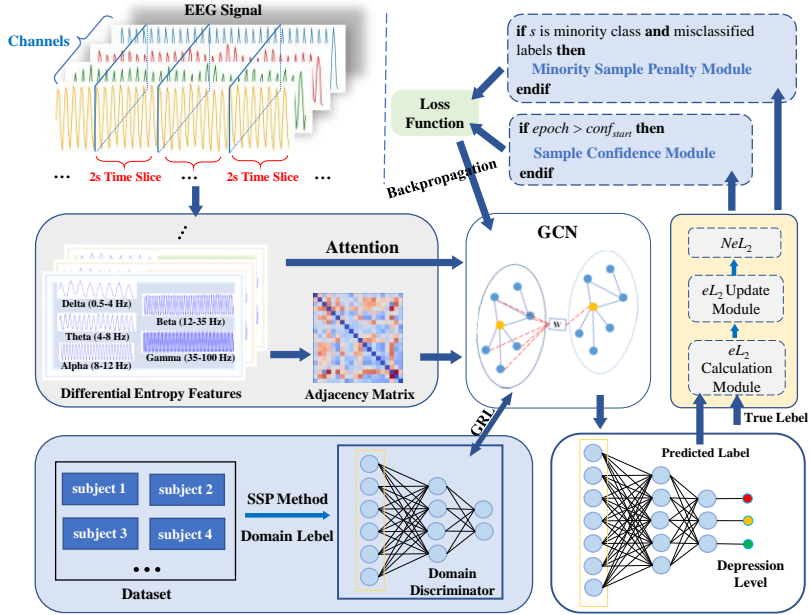

**Fig. 3.** Overall framework of the DepL-GCN model.

Studies have shown that in the early stages of training, the model remembers training data with clean and accurate labels before it remembers training data with noisy labels. The later the training time, the larger the error in weights updated due to mislabeling after the model converges to clean data. This phenomenon is intuitively demonstrated in Fig. 4. The model performs large-scale parameter updates in the early stages of training to quickly improve the fit for a large number of samples. As the training progresses, the model gradually turns to learning more complex samples, which often have unclear correspondence with labels and are therefore difficult to fit. In this process, the model may over-tune its parameters in order to force fit these samples, resulting in overfitting. Overfitting not only reduces the generalization ability of the model on the training set, but also affects the performance of the model on unknown data.

The learning rules of the deep learning model provide ideas for the model design in our study. Specifically, for the EEG-based depression-level recognition task, this study proposes a new sample confidence evaluation module for the model training process to reduce the degree of model learning for complex and inaccurate samples. Sample confidence evaluation is a technology for adaptively adjusting model prediction weights for different samples. During training, the model adjusts its learning rate according to the confidence level of the sample in order to focus more on those samples that the model predicts with more certainty. That is, after the model is trained to a certain stage, the sample confidence evaluation method proposed in this paper is introduced to appropriately reduce the learning of samples with low confidence. Next, the details of the sample confidence module are presented.

*1) $eL_2$ Norm and Update Strategy*

The calculation of sample confidence needs to be based on some indicators that reflect the difference between the current model's predicted value and the actual label, thereby providing guidance for the calculation of sample confidence. In this study, we use the $eL_2$ norm as the basis for calculating sample confidence. $eL_2$ can well reflect the gap between the model's predicted value and the true value. The calculation formula of the $eL_2$ norm is as follows:

$$eL_2 = \|\hat{\mathbf{y}} - \mathbf{y}\|_2, \quad (1)$$

where $\hat{\mathbf{y}}$ represents the label predicted by the model, and $\mathbf{y}$ is the true label of the sample.

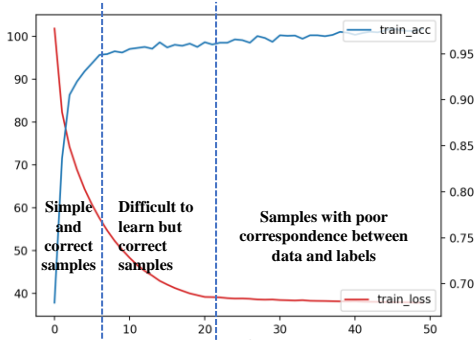

**Fig. 4.** Schematic diagram of the model training phase. The blue curve (train_acc) indicates the accuracy of the model, while the red curve (train_loss) shows the loss of the training data. The numbers on the left and right denote the training loss and accuracy, respectively. The dashed lines in the diagram are used to distinguish different training stages.

In addition, to reduce the model's sensitivity to abnormal $eL_2$ values and avoid misjudging occasional abnormal samples as low-confidence samples, this study optimizes the $eL_2$ update mechanism and introduces a smoothing strategy, that is, it does not directly use the subject's $eL_2$ in the current epoch for calculation, but comprehensively considers the $eL_2$ values of the previous epoch and the current epoch, and makes incremental adjustments based on the previous calculation results. As shown in Fig. 5. This strategy aims to achieve a robust update of the $eL_2$ value, thereby improving the overall generalization ability of the model. The calculation formula of $eL_2$ update strategy is as follows:

$$NeL_2 = LeL_2 - u\_rate * (LeL_2 - eL_2), \quad (2)$$

where $NeL_2$ refers to the $eL_2$ used for actual calculation based on the $eL_2$ of the current and previous epochs, $LeL_2$ represents the $eL_2$ of the sample in the previous epoch, and $u\_rate$ indicates the magnitude of updating this $eL_2$ based on the previous one.

*2) Sample Confidence Calculation*

The smaller the $eL_2$ value of a subject in the current training stage, the higher the confidence of the subject's sample. At the model level, it is considered that the data of the sample has a better correspondence with the label. It is desired that the model can learn as many samples as possible and establish a good correspondence between these data and labels. On the contrary, after certain epochs, if the $eL_2$ value of a subject is still large, the correspondence between the subject's EEG data and its depression level label may not be optimal, potentially due to noise or subjective interference from the patient. It is expected that after a certain period of training, the model will gradually abandon this part of the sample, rather than losing the prediction accuracy of other data in order to fit it. The calculation formula of the sample confidence value $val\_conf$ is as follows:

$$val\_conf = (1.0 - NeL_2 / \max(NeL_2)). \quad (3)$$

The maximum $eL_2$ value is established after a thorough analysis of extensive statistical data from both theoretical calculations and empirical tests, with the practical maximum being approximately 80% of the theoretical maximum. For enhanced precision, we adopt the statistical maximum as the default $\max(NeL_2)$. To prevent anomalies during training and to ensure that $val\_conf$ calculations are in the range of 0 to 1.0, an additional condition is incorporated. If the current $NeL_2$ exceeds the statistical maximum, the theoretical maximum is utilized in its place. The likelihood of scenario was assessed, and the MODMA dataset showed that this scenario occurred only twice during the comprehensive leave-one-out validation process, with a probability of less than 0.001% (2 occurrences / (52 subjects × 100 epochs × 53 experimental groups) ≈ 0.0007%). This rarity substantiates the accuracy and rationality of using the statistical maximum in our calculations.

*3) Sample Confidence Module Action Time*

During the model training process, the timing of introducing the sample confidence assessment method has a significant impact on model performance. The sample confidence evaluation method aims to optimize the model's continuous learning of learned samples, while gradually reducing the focus on difficult-to-learn samples. However, this strategy is not suitable for immediate implementation in the early stages of training. There are two main reasons:

First of all, the model will encounter two types of difficult samples: one is samples with correct labels, which are inherently hard to learn, and we expect the model to diligently master; The other is samples with a weak correlation between data and labels, which the model should minimally engage with. Ideally, the sample confidence evaluation should commence after the model has grasped the challenging yet correct samples, precluding engagement with poorly corresponding samples. Secondly, the delineation between these scenarios is ambiguous, complicating the precise timing for introducing sample confidence assessment. Typically, this requires tuning the hyperparameters through empirical exploration.

To address this, we postpone the introduction of sample confidence evaluation until the model has substantially learned from simpler samples, gradually integrating this mechanism to more precisely target and manage the more complex samples encountered later in training. This approach is designed to prevent performance deterioration due to overfitting.

Through an extensive experimental study, we have found that introducing the sample confidence evaluation between 40% and 60% of training completion is optimal for our model and hyperparameter configuration. We also meticulously examined various introduction timings to pinpoint the moment that best enhances the model's generalization and avoids overfitting. This strategy enables tailored treatment of simple and complex samples, as well as those with strong or weak data-label correspondence at different training stages, effectively balancing rapid learning of simple samples with fine-tuning of complex ones. Ultimately, this method bolsters the model's overall performance. The overall schematic of the sample confidence module is depicted in Fig. 6.

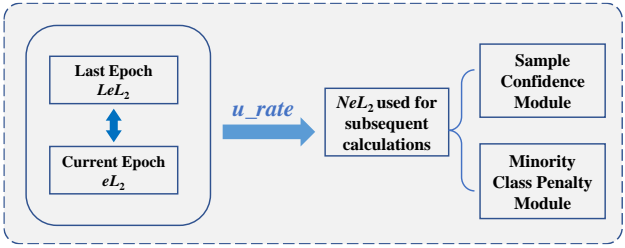

Fig. 5. Schematic diagram of *eL*₂ parameter update strategy.

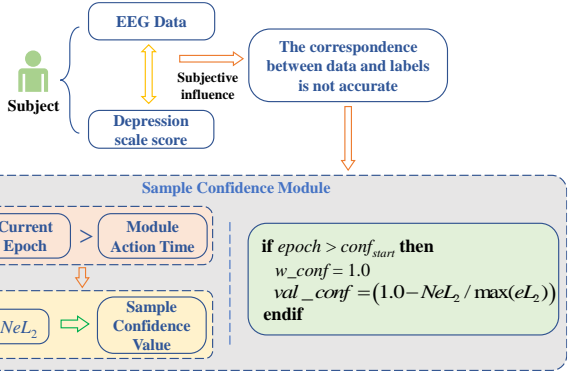

Fig. 6. Algorithm diagram of the sample confidence module.

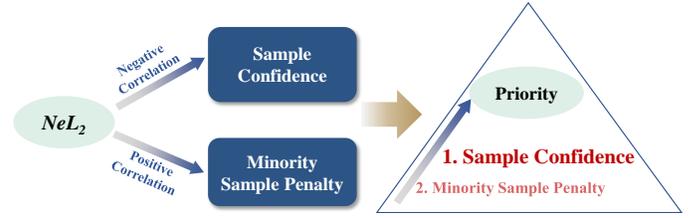

Fig. 7. Priorities of different modules.

### D. Minority Sample Penalty Module

The datasets we employed were initially designed for binary classification of depression, equitably representing both healthy subjects and those with depressive disorders. However, for the purpose of depression level recognition, it becomes necessary to further categorize patients with depression into different severity levels based on their scores on depression scales. This categorization inevitably leads to an imbalance in the number of samples across different depression severity categories, with the number of healthy subjects significantly exceeding that of any single depression severity category. Such imbalance may cause the model to develop a bias towards categories with a larger number of samples, neglecting those with fewer samples.

To address this issue, our study introduces a novel minority sample penalty module that adjusts the model's loss function based on the previously mentioned Normalized Euclidean Distance ($NeL_2$) to enhance the model's predictive capability for minority samples. The calculation method for the minority sample penalty coefficient, denoted as *val_pen*, is as follows:

$$val\_pen = NeL_2 / \max(eL_2). \quad (4)$$

Specifically, when the model incorrectly predicts a sample during training, and that sample belongs to a category with a smaller number of samples, a penalty value *val_pen* is computed and incorporated into the sample's loss function. The intention is to increase the weight of minority class samples within the loss function, thereby drawing the model's attention to these samples. The penalty coefficient for minority samples is not static but positively correlated with the $NeL_2$ value. In other words, the larger the prediction error of the model for a sample, the greater the associated penalty coefficient.

In this study, based on the number of subjects in each depression severity level obtained from TABLE II, for the MODMA dataset, the categories of mild depression (5 subjects) and moderate depression (3 subjects) are classified as minority samples. For the PRED+CT dataset, the categories of major depression (5 subjects), moderate depression (23 subjects), and mild depression (14 subjects) are considered as minority samples.

By introducing this $NeL_2$ based minority sample penalty coefficient, the model will place greater emphasis on minority class samples that are incorrectly predicted during training, thereby improving the model's predictive ability for these classes. This approach helps to balance the differences in sample numbers between classes, enhances the generalization performance of the model, and reduces the neglect of minority samples.

### E. Loss Function

During the training process, the calculation method of the base class loss $Loss_{class}$ and domain loss $Loss_{Domain}$ for each subject is as follows:

$$Loss_{class} = -\frac{1}{R}\sum_{r=1}^{R}\sum_{n=1}^{N}\mathbf{y}_{r,n} \log \hat{\mathbf{y}}_{r,n}, \quad (5)$$

$$Loss_{domain} = -\frac{1}{R}\sum_{r=1}^{R}\sum_{dn=1}^{DN}\mathbf{d}_{r,dn} \log \hat{\mathbf{d}}_{r,dn}, \quad (6)$$

where $R$ represents the number of samples, and $N$ denotes the number of classes in the classification task. For the MODMA dataset, $N=5$, and for the PRED+CT dataset, $N=4$. $\mathbf{y}_{r,n}$ signifies the true label of the $r$-th sample belonging to the $n$-th class, while $\hat{\mathbf{y}}_{h,n}$ indicates the predicted class label obtained through the model. $DN$ refers to the number of domain label categories in the SSPA-GCN model. $\mathbf{d}_{r,dn}$ is the true domain label of the $r$-th sample in the $dn$-th domain, and $\hat{\mathbf{d}}_{r,dn}$ stands for the predicted domain label calculated by the domain discriminator.

In order to integrate the proposed sample confidence module and minority sample category penalty coefficient module into the model, we implement this mechanism by optimizing the loss function. For subject $i$, the weight coefficients of these two modules in the loss function can be expressed as:

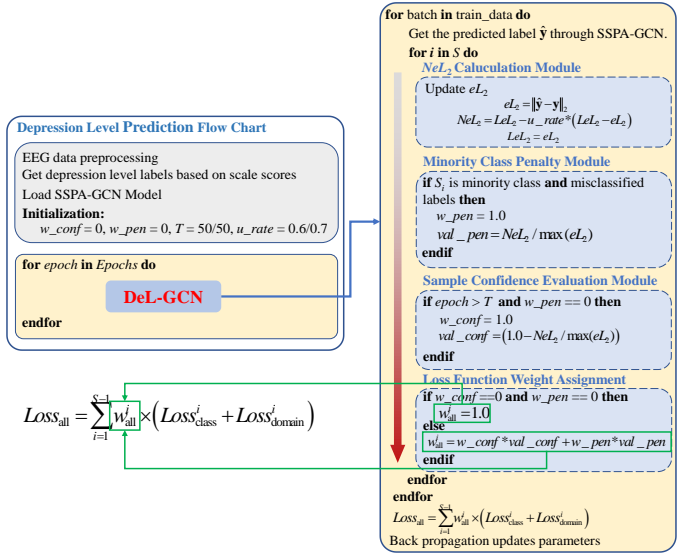

**Fig. 8.** Training process of the depression-level prediction algorithm.

$$w_{all}^j = w\_conf * val\_conf_i + w\_pen * val\_pen_i. \quad (7)$$

Among them, $w\_conf$ and $w\_pen$ are the weighted coefficients of the sample confidence module and the minority sample penalty module, respectively, which are mainly used to control whether the different modules in the model are enabled. It is a Boolean value and can only take the value of 0 or 1. $val\_conf$ and $val\_pen$ denote the specific values calculated by the two modules. $w_{all}^j$ indicates the weighted sum of the loss values of subject $i$, which is added to the loss function of the model as the total loss weight of subject $i$.

Therefore, in every training epoch, the total loss function of the DepL-GCN model can be expressed as:

$$Loss_{all} = \sum_{i=1}^{S} w_{all}^j \times \left( Loss_{class}^i + Loss_{domain}^i \right). \quad (8)$$

In the sample confidence assessment mechanism, we have observed a negative correlation between the sample confidence coefficient and the $eL_2$ value. Conversely, in the minority sample penalty mechanism, the penalty coefficient exhibits a positive correlation with $eL2$. If a sample meets the triggering conditions for both modules at a specific time point, their combined effects, which rely on independent $eL_2$ calculations, may neutralize each other, thus failing to achieve the desired optimization goal.

To mitigate this phenomenon, we assign different priorities to the sample confidence assessment and the sample penalty mechanism during the model training process, as shown in Fig. 7. Given our preference for maintaining the model's learning ability for minority samples, we grant a higher priority to the minority sample penalty mechanism. Therefore, when both modules are triggered, the sample confidence assessment module will be suppressed within the current training cycle, with $w\_conf$ being set to 0. This ensures that the minority sample penalty mechanism can exert its proper function.

Furthermore, in the early stages of training, samples that do not belong to minority samples or minority class samples that are predicted correctly will not trigger either the sample confidence module or the minority sample penalty module, meaning both $w\_conf$ and $w\_pen$ are 0. In this case, the weight will be programmatically set to 1.0 to ensure that the model's training process can proceed normally.

In summary, the complete training process of depression level prediction is shown in Fig. 8.

## IV. EXPERIMENT AND DISCUSSION

### A. Datasets

The MODMA dataset [16], introduced by Lanzhou University's UAIS lab in 2020, serves psychological disorder analysis. It encompasses EEG data from 53 clinically diagnosed depressive patients and 55 healthy counterparts selected by psychiatrists. The dataset's 128-channel EEG signals, captured during rest and a dot probe task with HydroCel Geodesic Sensor Net and Net Station (v4.5.4), were used in this study. We focused on 128-channel resting-state EEG from 53 subjects (24 patients, 29 controls). Prior to the study, subjects completed PHQ-9, GAD-7, and psychiatric evaluations. The dataset underwent preprocessing and quality checks.

The PRED+CT [17] initiative aims to facilitate large-scale data mining for neurological and psychiatric biomarker development. Its depression dataset records EEG signals from depressive patients screened via the BDI and clinical interviews. Participants with a BDI score under 7 were deemed healthy, while those above were classified as patients. EEG data, collected at 500 Hz through 64 Ag/AgCl electrodes, were preprocessed by the data providers.

### B. Experiment Settings

In the experiment, we used all subjects in the MODMA and PRED+CT datasets with the required data length (5 minutes), and divided the subjects in the dataset into depression levels according to the depression-level classification criteria provided by the PHQ-9 and BDI questionnaires. The classification criteria are shown in TABLE I. The detailed information of the dataset and the number of subjects at different depression levels are detailed in TABLE II.

During the experiments, the leave-one-out cross-validation method was used to comprehensively test the model. The experimental results were statistically analyzed using the subject as the basic calculation unit. All experiments were conducted under uniform hardware conditions to ensure the consistency and comparability of the results. The specific hardware configuration is as follows: the operating system is Ubuntu 22.04, equipped with four RTX 4090D GPU, and the main software environment includes PyTorch 1.12.0 and Numpy 1.24.0.

We used macro-average and micro-average methods as evaluation indicators of the model [29]. The macro-average method can calculate separate indicators for each category and provide insights into the performance of each category. The micro-average method requires global statistical dataset classification results to calculate the overall performance indicators.

The macro-average precision, denoted as $Pre_{macro}$, can be expressed as:

TABLE I
DEPRESSION LEVEL CLASSIFICATION CRITERIA

| PHQ-9 score | BDI score | Depression level |
|---|---|---|
| 20-27 | 29-63 | Major |
| 15-19 | / | Moderate to Major |
| 10-14 | 20-28 | Moderate |
| 5-9 | 14-19 | Mild |
| 0-4 | 0-13 | Normal |

TABLE II
DETAILED INFORMATION ABOUT SUBJECTS IN THE DATASETS

| | | MODMA | PRED+CT |
|---|---|---|---|
| Number of all participants | | 53 | 118 |
| Depression rating scale | | PHQ-9 | BDI |
| Number of subjects in each depression level | Major | 8 | 5 |
| | Moderate to Major | 13 | / |
| | Moderate | 3 | 23 |
| | Mild | 5 | 14 |
| | Normal | 24 | 76 |
| Gender (M/F) | | 33/20 | 71/47 |
| Number of channels | | 128 | 64 |
| Sampling rate | | 250 Hz | 500 Hz |

$$Pre_{macro} = \frac{1}{N}\sum_{n=1}^{N} TP_n/(TP_n + FP_n), \quad (9)$$

where $TP_n$ represents the number of samples correctly predicted as belonging to the $n$-th category, while $FP_n$ denotes the number of samples incorrectly predicted to belong to the $n$-th category. For the MODMA dataset, $N=5$, and for the PRED+CT dataset, $N=4$.

The macro-average recall, denoted as $Rec_{macro}$, can be expressed as:

$$Rec_{macro} = \frac{1}{N}\sum_{n=1}^{N} TP_n/(TP_n + FN_n), \quad (10)$$

where $FN_n$ represents the number of samples that belong to the $n$-th category but are incorrectly predicted as not belonging to that category.

The calculation formula for the macro-average $F1$ score is as follows:

$$F1_{macro} = 2\times\frac{Pre_{macro}\times Rec_{macro}}{Pre_{macro} + Rec_{macro}}. \quad (11)$$

The calculation for micro-average precision, micro-average recall, and micro-average $F1$ score can be represented as follows:

$$Pre_{micro} = \frac{\sum_{i=n}^{N} TP_n}{\sum_{n=1}^{N} TP_n + \sum_{n=1}^{N} FP_n}, \quad (12)$$

$$Rec_{micro} = \frac{\sum_{i=n}^{N} TP_n}{\sum_{n=1}^{N} TP_n + \sum_{n=1}^{N} FN_n}, \quad (13)$$

$$F1_{micro} = 2\times\frac{Pre_{micro}\times Rec_{micro}}{Pre_{micro} + Rec_{micro}}. \quad (14)$$

In multi-classification problems, $Pre_{micro}$, $Rec_{micro}$ and $F1$ score are usually equal because they are all calculated based on global $TP$, $FP$, and $FN$. Therefore, in the process of statistical micro-average indicators, we only calculate the micro-average precision $Pre_{micro}$ as a representative.

*C. Ablation Experiment*

In order to verify the effectiveness of each module in the model proposed in this paper, a series of different models were designed and ablation experiments were performed. The purpose of the ablation experiment is to evaluate the contribution of each module in the model to the final performance one by one. The specific sub-models are described as follows:

(1) Model A (base model): The SSPA-GCN model is used as the base model. This model does not contain any modules proposed in this study and serves as the control group for subsequent experiments.

(2) Model B (+ sample confidence module): Based on model A, only the sample confidence module is added. This module aims to evaluate the model's confidence level for each sample so that samples with low confidence can be appropriately discarded during training.

(3) Model C (+ minority sample penalty coefficient module): Based on Model A, only the minority sample penalty coefficient module is introduced, which is used to deal with the category imbalance problem in multi-classification tasks.

(4) Model D (DepL-GCN): Based on Model A, the sample confidence module and the minority sample penalty module are added at the same time.

For the models discussed, extensive experiments were conducted on the MODMA and PRED+CT datasets, and the performance metrics were statistically analyzed. TABLE III present the performance statistics of the ablation experiments on these datasets, while

**Fig. 9** shows the corresponding confusion matrices.

First, we analyze the performance of different models on the MODMA dataset. Model A, directly applied to the multi-classification task, achieved a classification accuracy of 71.70%. Notably, without the minority sample penalty module, its recall rates for moderate and mild depression were only 33.33% and 20.00%, respectively. The confusion matrix in attention to minority sample categories. Although its overall accuracy was only 3.77% higher than Model A, its recall rates for moderate and mild depression were improved by 33.34% and 40.00%, respectively. This demonstrates the effectiveness of the minority sample penalty module in handling imbalanced datasets, with the $NeL_2$-based penalty calculation proving reasonable. Finally, Model D, combining both the sample confidence and minority sample penalty modules, achieved an accuracy of 81.13%, a 9.43% increase over Model A. This surpasses the performance improvement of each module alone, maintaining good results for minority sample categories. In summary, the experiments on the MODMA dataset validate the effectiveness of the proposed modules.

Fig. **9** reveals that Model A correctly classified only two subjects in the minority sample category, often misclassifying unclassifiable samples as healthy due to the large number of healthy subjects. Consequently, the recall rate for the healthy category reached 95.83%. Model B, incorporating the sample confidence module, improved its classification accuracy to 79.25%, a 7.55% increase over Model A. This improvement is attributed to the module's strategy of gradually discarding samples with poor data-label correspondence, enhancing the model's generalization. However, Model B's performance on minority sample categories remained unchanged, with a tendency to classify all mis-predicted samples as healthy. Model C, with the minority sample penalty module, showed significant

TABLE III
EXPERIMENTAL RESULTS OF ABLATION EXPERIMENTS

| Dataset | | MODMA | | | | | PRED+CT | | | | |
|---|---|---|---|---|---|---|---|---|---|---|---|
| Model | Depression Level | Acc (%) | Pre (%) | Rec (%) | F1 (%) | Pre$_{micro}$ (%) | Acc (%) | Pre (%) | Rec (%) | F1 (%) | Pre$_{micro}$ (%) |
| Model A | Major | 71.70 | 62.50 | 62.50 | 62.50 | 71.70 | 70.34 | 0.00 | 0.00 | 0.00 | 70.34 |
| | Moderate to Major | | 100.00 | 61.54 | 76.19 | | | - | - | - | |
| | Moderate | | 100.00 | 33.33 | 50.00 | | | 72.73 | 34.78 | 47.06 | |
| | Mild | | 100.00 | 20.00 | 33.33 | | | 43.75 | 50.00 | 46.67 | |
| | Normal | | 65.71 | 95.83 | 77.97 | | | 75.56 | 89.47 | 81.93 | |
| | **Macro Average** | | 85.64 | 54.64 | 60.00 | | | 48.01 | 43.56 | 43.91 | |
| Model B | Major | 79.25 | 85.71 | 75.00 | 80.00 | 79.25 | 75.42 | 0.00 | 0.00 | 0.00 | 75.42 |
| | Moderate to Major | | 90.91 | 76.92 | 83.33 | | | - | - | - | |
| | Moderate | | 100.00 | 33.33 | 50.00 | | | 73.33 | 47.83 | 57.89 | |
| | Mild | | 100.00 | 20.00 | 33.33 | | | 57.14 | 57.14 | 57.14 | |
| | Normal | | 72.73 | 100.00 | 84.21 | | | 79.55 | 92.11 | 85.37 | |
| | **Macro Average** | | 89.87 | 61.05 | 66.18 | | | 52.51 | 49.27 | 50.10 | |
| Model C | Major | 75.47 | 75.00 | 75.00 | 75.00 | 75.47 | 78.81 | 40.00 | 40.00 | 40.00 | 78.81 |
| | Moderate to Major | | 87.50 | 53.85 | 66.67 | | | - | - | - | |
| | Moderate | | 66.67 | 66.67 | 66.67 | | | 76.19 | 69.57 | 72.73 | |
| | Mild | | 75.00 | 60.00 | 66.67 | | | 57.89 | 78.57 | 66.67 | |
| | Normal | | 73.33 | 91.67 | 81.48 | | | 87.67 | 84.21 | 85.91 | |
| | **Macro Average** | | 75.50 | 69.44 | 71.30 | | | 65.44 | 68.09 | 66.32 | |
| Model D | Major | 81.13 | 85.71 | 75.00 | 80.00 | 81.13 | 81.36 | 33.33 | 40.00 | 36.36 | 81.36 |
| | Moderate to Major | | 90.00 | 69.23 | 78.26 | | | - | - | - | |
| | Moderate | | 100.00 | 100.00 | 100.00 | | | 74.07 | 86.96 | 80.00 | |
| | Mild | | 100.00 | 40.00 | 57.14 | | | 65.00 | 92.86 | 76.47 | |
| | Normal | | 74.19 | 95.83 | 83.64 | | | 93.85 | 80.26 | 86.52 | |
| | **Macro Average** | | 89.98 | 76.01 | 79.81 | | | 66.56 | 75.02 | 69.84 | |

For the PRED+CT dataset, the results mirror the overall trend observed in the MODMA dataset. Model A's accuracy of 70.34% is unsatisfactory, as simply predicting all samples as healthy would yield 64.41% accuracy. Its recall rate for the healthy category was 89.47%, but it failed to classify any major depression cases, achieving 0% accuracy. The confusion matrix in

Fig. 9 shows that, aside from the healthy category, Model A correctly classified only 15 subjects, highlighting the issue of sample imbalance. Model B, with the sample confidence module, improved its accuracy to 75.42%, a 5.08% increase over Model A. This indicates that the module effectively reduced attention to samples with low data-label correspondence, enhancing classification performance. However, it still struggled with the major depression category, maintaining a 0% recall rate. Model C, incorporating the minority sample penalty module, increased its overall accuracy by 8.47% compared to

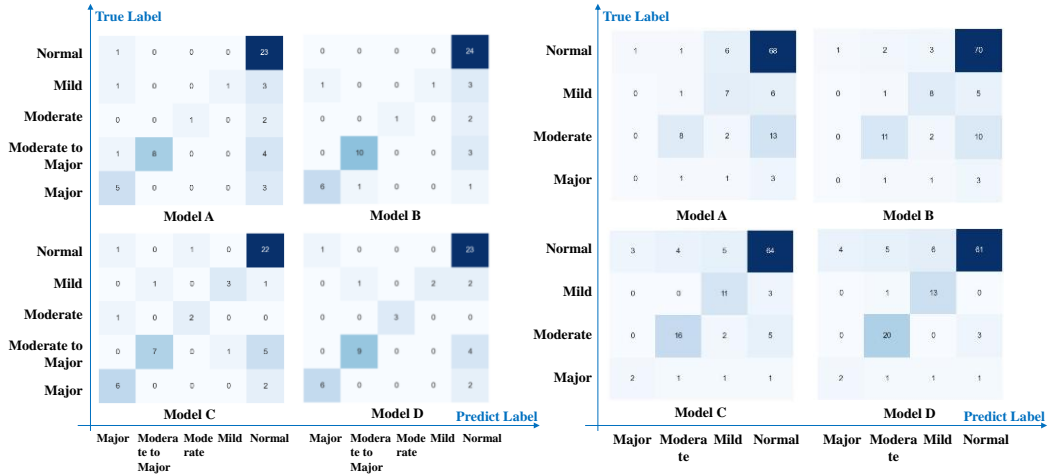

Fig. 9. Confusion matrix of the experiments, with the left and right sides representing the MODMA and PRED+CT datasets, respectively.

Model A. Its recall rates for major, moderate, and mild depression reached 40%, 69.57%, and 78.57%, respectively. Notably, Model C successfully identified some major depression samples, outperforming Models A and B. Finally, Model D, combining both modules, achieved an accuracy of 81.36%, a 11.02% increase over Model A. This improvement exceeds that of each module alone.

The ablation experiment results demonstrate that the sample confidence and minority sample penalty modules effectively enhance model performance in multi-classification depression level tasks, with combined use yielding greater improvements than either module alone.

### D. Comparison with the SOTA Models

Due to the limited number of studies on depression level recognition using EEG signals and the diversity in datasets and validation methods, direct and fair comparisons of experimental results are challenging. To evaluate the performance of our proposed DepL-GCN model against existing models, we replicated several state-of-the-art EEG signal processing models. We maintained consistent settings across these models during data processing and cross-validation. The results of our replicated experiments are based on the MODMA and PRED+CT datasets. Below, we briefly introduce these models:

(1) CNN [30]: Convolutional neural networks, which include structures such as convolutional layers, pooling layers, and fully connected layers.

(2) EEGNet [31]: A neural network model specifically designed for EEG classification. It includes components such as convolutional layers, depth-wise separable convolutional layers, and fully connected layers.

(3) GICN [32]: An improved GCN network that applies a weighted matrix on the adjacency matrix of graph convolution to redefine and calculate the connectivity between different EEG channels.

(4) AMGCN-L [33]: A GCN+LSTM architecture that uses a multi-time window graph construction method, designed for EEG-based depression recognition tasks.

(5) LSDD_EEGNet [34]: A model based on the CNN+LSTM architecture, with a gradient reversal layer applied after the LSTM to implement domain adversarial training, obtaining subject-invariant features to enhance the model's generalization capability.

To ensure the transparency of the model replication process and the verifiability of the experimental results, we conducted strict parameter configuration for all comparative models, ensuring the optimal performance of each model. The final experimental results are shown in TABLE IV and the confusion matrix is illustrated in Fig. 10.

The experimental results in TABLE IV highlight the significant advantages of the proposed DepL-GCN model in depression level recognition. On the MODMA dataset, DepL-GCN outperforms other models, with AMGCN-L achieving the next highest classification accuracy at 73.58%, 7.55% lower than DepL-GCN. While AMGCN-L is not designed for depression level classification, it still shows potential in this task. Other models, including EEGNet and LSDD_EEGNet, demonstrate lower accuracies of 60.38% and 67.92%, respectively, often misclassifying minority sample categories as healthy. The CNN model has the lowest accuracy of 50.94%, barely above random guess of 45.28%. The confusion matrix reveals that CNN correctly classified only 4 subjects in moderate and moderate to major categories, with a 0% recognition rate for minority sample categories of moderate and mild depression, indicating near failure. These models are unsuitable for EEG-based depression level without specific improvements. DepL-GCN, with its specialized improvements, shows superior performance.

On the PRED+CT dataset, DepL-GCN again leads with the best results. LSDD_EEGNet follows with a classification accuracy of 72.03%. GICN and CNN models report accuracies of 63.56% and 60.17%, respectively, close to random prediction or classifying all subjects as healthy (64.41%). Except for DepL-GCN, the macro-average performance metrics of all models are mostly below 50%. This is due to the few sample sizes outside of the healthy category, which are difficult for these models to identify correctly, leading to poor performance metrics or even 0% indicators. Thus, the macro-average metric, calculated by averaging performance across categories, reflects a reduced model performance. This underscores the necessity of designing models specifically for depression-level classification tasks.

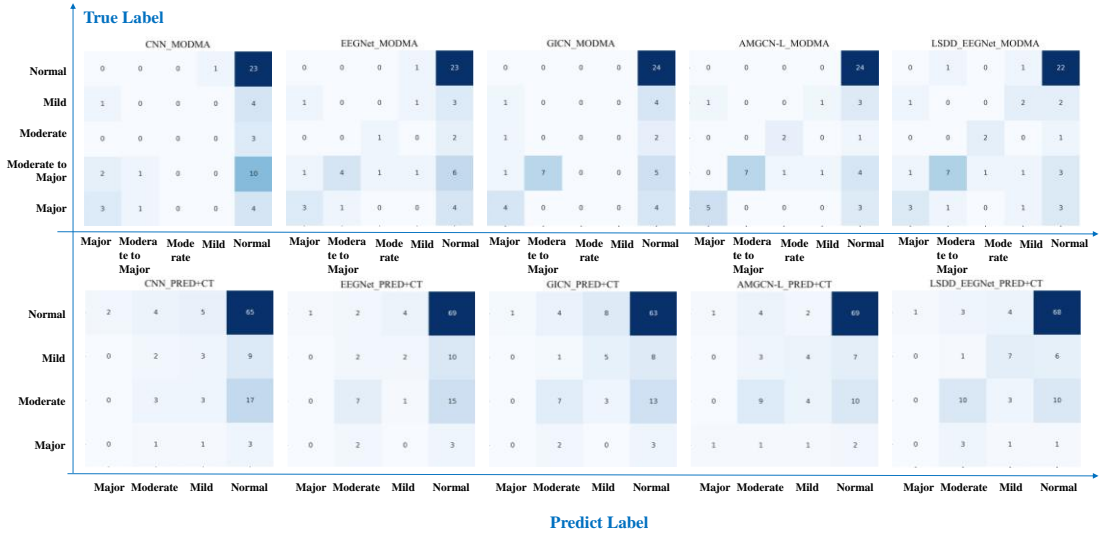

**Fig. 10.** Confusion matrix of comparative experimental results.

TABLE IV
PERFORMANCE COMPARISON OF DIFFERENT MODELS

| Datasets | Model | Acc (%) | Pre$_{macro}$ (%) | Rec$_{macro}$ (%) | F1$_{macro}$ (%) | Pre$_{micro}$ (%) |
|---|---|---|---|---|---|---|
| MODMA | CNN | 50.94 | 30.45 | 28.21 | 24.77 | 50.94 |
| | EEGNet | 60.38 | 56.67 | 43.49 | 45.96 | 60.38 |
| | GICN | 66.04 | 43.74 | 40.77 | 39.90 | 66.04 |
| | AMGCN-L | 73.58 | 73.31 | 60.60 | 63.60 | 73.58 |
| | LSDD_EEGNet | 67.92 | 63.08 | 57.94 | 59.29 | 67.92 |
| | DepL-GCN | 81.13 | 89.98 | 76.01 | 79.81 | 81.13 |
| PRED+CT | CNN | 60.17 | 31.04 | 30.00 | 29.43 | 60.17 |
| | EEGNet | 66.10 | 38.29 | 33.88 | 34.43 | 66.10 |
| | GICN | 63.56 | 38.42 | 37.26 | 37.12 | 63.56 |
| | AMGCN-L | 70.34 | 54.43 | 44.62 | 47.43 | 70.34 |
| | LSDD_EEGNet | 72.03 | 46.37 | 45.74 | 45.69 | 72.03 |
| | DepL-GCN | 81.36 | 66.56 | 75.02 | 69.84 | 81.36 |

*E. Hyperparameter Experiment*

*1) eL$_2$ Update Rate*

The update rate of the $eL_2$ parameter $u\_rate$, is a key hyperparameter that influences the model's classification performance and cannot be theoretically calculated. Therefore, to thoroughly understand the specific impact of $u\_rate$ on model performance, we conducted a series of experiments, and the results are displayed in Fig. 11.

For the MODMA dataset, the optimal $u\_rate$ is from 0.5 to 0.6, while for the PRED+CT dataset, it is from 0.7 to 0.9. Within these ranges, the model balances sensitivity to data features and robustness against noise.

By controlling the parameter $u\_rate$, we observed that increasing update frequency initially improves, then decreases model classification accuracy. When $u\_rate$ is low, increasing it enhances performance, likely because higher frequency helps the model better adapt to changes in $eL_2$ during training. However, beyond a critical value, performance stabilizes or declines due to increased sensitivity to noise and outliers.

*2) Sample Confidence Module Start Epoch*

The action time of the sample confidence evaluation module affects the model's learning process for samples of varying difficulty and data-label correspondence, thus impacting model performance. Since the start action time $conf_{start}$, cannot be theoretically calculated, we conducted experiments with multiple $conf_{start}$ times (100 epochs total, with each 10 epochs as a node) and statistically analyzed the results, as shown in Fig. 12.

The experimental results indicate the best timing for introducing the sample confidence assessment module on different datasets. For the MODMA dataset, it should be introduced between the 40th and 50th training epochs, while for the PRED+CT dataset, it is best to start between the 50th and 70th epochs. At these points, the model has effectively learned simpler samples and needs additional help to manage complex or ambiguous ones. Introducing the module too early may hinder learning of basic or complex samples, leading to underfitting. Conversely, introducing it too late may cause the model to overfit to poorly correlated samples, harming its test set performance. Thus, choosing the right timing for the sample confidence assessment module is key to optimizing model performance and preventing overfitting.

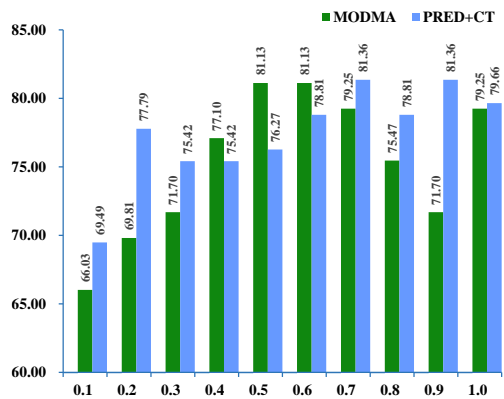

**Fig. 11.** Experimental results of *u_rate* hyperparameter.

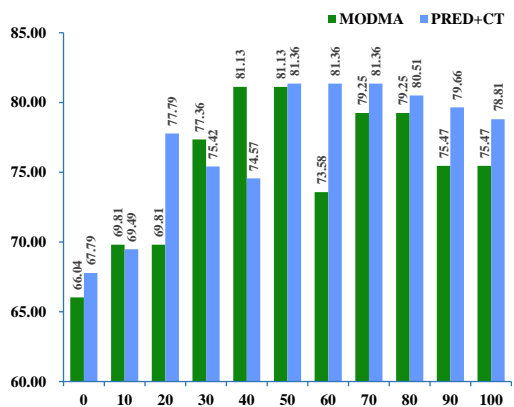

**Fig. 12.** Experimental results of sample confidence module action time hyperparameter.

*F. Discussion on Depression Level Classification Task Based on Regression Model*

In the depression level identification task, in addition to the multi-classification model mentioned above, the regression model is also widely used in the depression level prediction task [18], [24]. By predicting the score of the subject's depression level, rather than directly predicting the depression level through classification, it can provide higher accuracy and fine-grained assessment results. However, at the same time, it has higher requirements on the rationality of the model design and the accuracy of the data labels. In this section, we trained a depression-level detection model based on regression tasks by slightly adjusting the DepL-GCN model structure and hyperparameters during the training process. We briefly discuss the depression-level prediction model and its performance based on regression tasks.

Similarly, we conducted experiments on both the MODMA and PRED+CT datasets, and the final experimental results are shown in TABLE V.

We conducted depression score prediction based on EEG and achieved the optimal MAE of 6.32 on the MODMA dataset and 9.32 on the PRED+CT dataset. These errors have the potential to be further reduced through the rational design of models or the improvement of data quality. Additionally, some issues have been identified through experimental observations, such as the asymmetric risk [35] in the predictions of regression models. This leads to a situation where, even with a small MAE, the predicted scores may not accurately reflect the patients' depression levels. If the depression levels are directly classified based on the predicted regression scores, the classification accuracy will be below 50%. For subjects whose depression-level scores are near the category boundaries, regression models often fail to make accurate predictions.

TABLE V
EXPERIMENTAL RESULTS OF DEPRESSION-LEVEL RECOGNITION MODEL BASED ON REGRESSION TASK

| Dataset | MAE | RMSE |
| --- | --- | --- |
| MODMA | 6.32 | 7.54 |
| PRED+CT | 9.32 | 12.14 |

Overall, although regression models offer a new perspective for predicting depression levels, there is still a long way to go before they can be practically applied and used to guide doctors in their judgments. Future research needs to delve deeper into the design of model structures, the enhancement of data quality, and the clinical interpretation of predicted scores.

## V. CONCLUSION

We completed the task of identifying subjects' depression levels on two publicly available EEG datasets containing depressive disorders and healthy subjects. In order to solve the problem of lax correspondence between data and labels in the depression-level identification process and the imbalance of the number of samples between different categories, we innovatively proposed a sample confidence module and a minority sample penalty module based on the second norm of errors. Extensive ablation experiments and comparative experiments on the two datasets demonstrate that our proposed modules for the depression-level recognition task are effective. In addition, we explored the regression task-based depression-level identification model, which provides a new perspective by predicting continuous depression-level scores instead of direct classification results.

Under the current circumstances, there are still some limitations in our work. First, the calculation of sample confidence can explore more accurate and adaptable methods, such as considering the integration of more sample feature information, rather than relying solely on $NeL_2$. Second, future research should strengthen the application of the model in clinical environments, addressing potential issues that may arise in practical applications.